\documentclass[prl,aps, twocolumn, showpacs,amsmath,amssymb,floatfix,superscriptaddress,citeautoscript]{revtex4-1}
\usepackage{graphicx}
\usepackage{dcolumn}
\usepackage{bm}
\usepackage{amsmath}
\usepackage{amssymb}
\usepackage{amssymb, verbatim}
\usepackage[normal]{subfigure}

\usepackage{hyperref}
\hypersetup{
        colorlinks=true,
}

\begin{document}

\title{Transport through a Majorana island: strong tunneling regime }
\author{Roman M. Lutchyn}
\affiliation{Station Q, Microsoft Research, Santa Barbara, California 93106-6105, USA}
\author{Leonid I. Glazman}
\affiliation{Department of Physics, Yale University, New Haven, CT 06520, USA}

\date{\today}

\begin{abstract}
In the presence of Rashba spin-orbit coupling, magnetic field can drive a proximitized nanowire into a topological superconducting phase~\cite{Lutchyn10,Oreg10}. We study transport properties of such nanowires in the Coulomb blockade regime. The associated with the topological superconductivity Majorana modes significantly modify transport and lead to single-electron coherent transmission through the nanowire - a non-local signature of topological superconductivity. In this work, we focus on the case of strong hybridization of the Majorana modes with the normal leads.
%consider strong tunneling regime where Majorana modes are strongly hybridized with the normal leads.
The induced by hybridization broadening of the Majorana zero-energy states competes with the charging energy, leading to a considerable modification of the Coulomb blockade in a nanowire contacted by two normal leads. We evaluate the two-terminal conductance as a function of the gate voltage, junctions transmission coefficients, the geometric capacitance of and the induced superconducting gap in the nanowire.
\end{abstract}

\maketitle

%{\it Introduction.}
Topological superconductors provide a promising platform for fault-tolerant quantum computation~\cite{Nayak08, Beenakker13a, Alicea12a, Leijnse12, Stanescu13b, DasSarma15}. These exotic electronic phases of matter are predicted to host defects binding Majorana zero-energy modes which obey non-Abelian braiding statistics~\cite{Read00, Ivanov, Alicea11}. Theory predicts that Majorana zero modes may be realized at the ends of proximitized nanowires~\cite{Sau10, Lutchyn10, Oreg10}, and there is mounting experimental evidence for their existence in these systems~\cite{Mourik2012,Das2012,Deng2012,Fink2012,Churchill2013, Deng2014,Higginbotham15,albrecht2015,HaoZhang16, Deng16}.

\begin{figure}
\begin{centering}
\includegraphics[width=\columnwidth]{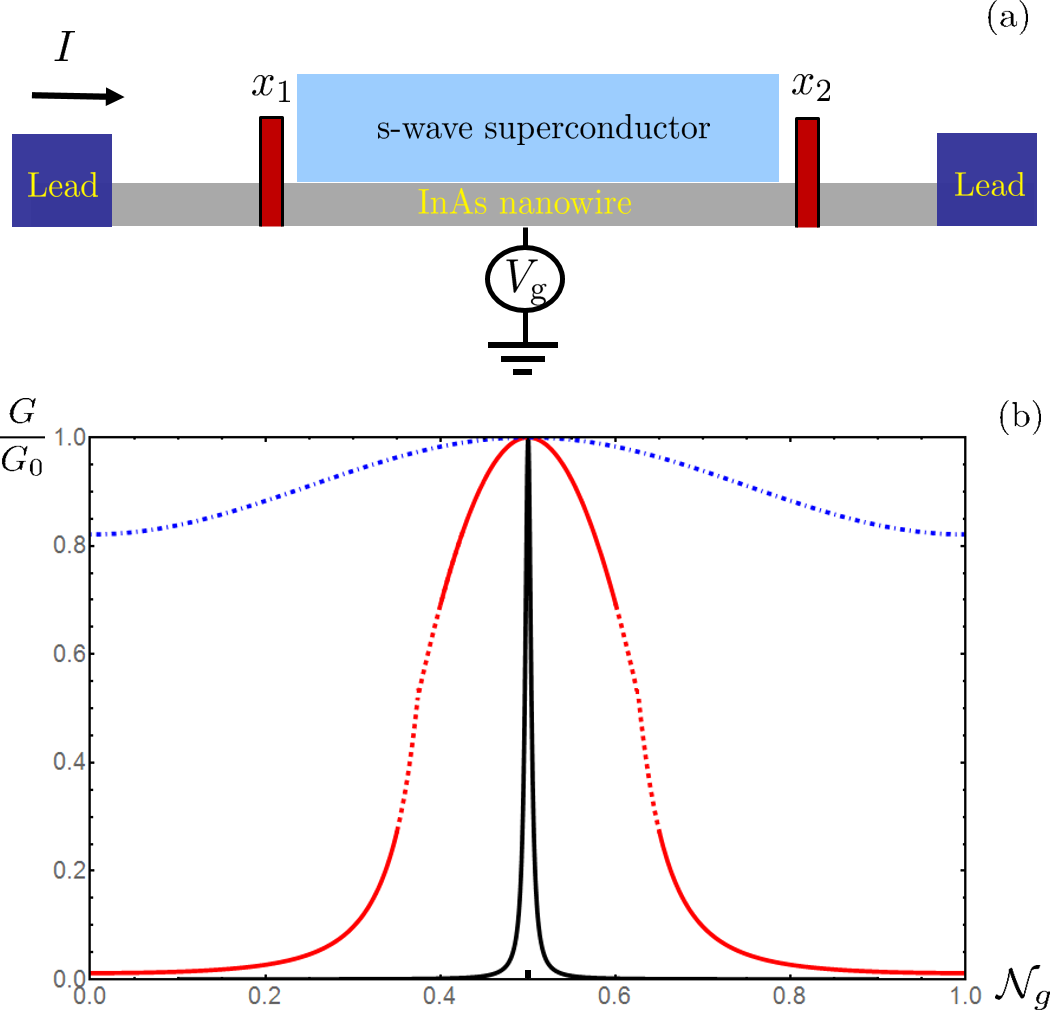}
\par\end{centering}
\caption{(Color online) Panel (a): Schematic plot of the device. Panel (b): Conductance $G$ as a function of the dimensionless gate voltage ${\mathcal N}_g$. In a symmetric device, $g_1=g_2\equiv g$, conductance reaches $G_0$.
Solid (black) curve: Coulomb blockade peak at $g\ll 1$ is a Lorentzian~\cite{Heck16} of width given by Eq.~(\ref{eq:width3}). Dashed/solid (red) curve: $G({\cal N}_g)$ at intermediate values of $g$ such that $\Delta_P/E_C\ll 1-g\ll 1$; see Eq.~(\ref{eq:width1}) for the width of the maximum and Eq.~(\ref{eq:width2}) for the crossover to the weak-tunneling limit. Dot-dashed (blue) curve: $G({\cal N}_g)$ of a symmetric device in the strong-tunneling limit, $E_C(1-g)\ll \Delta_P$. Conductance approaches the unitary limit, exhibiting weak ${\mathcal N}_g$-dependent oscillations, see Eq.~(\ref{eq:conductance_weak}).}
\label{fig:device}
\end{figure}

Most of the proposals for Majorana-based topological quantum computation involve mesoscopic islands with a sizable charging energy which contain two or more Majorana modes (Majorana islands)~\cite{Hassler10, Bonderson11b,Sau11a, Hyart13, Aasen16, Landau16,Plugge16a, Plugge16, Karzig16}.  Therefore, it is important to understand the interplay of topological degrees of freedom and charging energy of these islands. Another motivation comes from the recent experiment by Albrecht {\it et al.}~\cite{albrecht2015} investigating the dependence of two-terminal conductance through a Majorana island in the Coulomb blockade regime, see Fig.~\ref{fig:device}a for the device layout.
 The existing theory~\cite{Fu10, Heck16} allows one to evaluate the conductance of a Majorana island in the weak tunneling regime, $g_i\ll 1$, using resonant level approximation (here $g_i$ is the dimensionless normal-state  conductance of $i$-th junction, $G_i=g_iG_0$, and $G_0=e^2/h$ is the conductance quantum for spin-polarized electrons). In that approximation, only the resonant level comprised of the two degenerate ground states of the island is involved in the formation of narrow Coulomb blockade conductance peaks, see Fig.~\ref{fig:device}b.

The resonant-level approximation, however, is inapplicable to the strong-tunneling regime, corresponding to one or both junctions approaching the reflectionless limit (i.e. $1-g_i \ll 1$). The width of the  broadened resonant level then becomes comparable to the topological gap $\Delta_P$. Under this condition, the quasi-continuum of excited states with energies above $\Delta_P$ also contributes to the electron transport across the island. The problem at hand is rather non-trivial. A similar setting in the absence of superconductivity and in the limit of zero spacing between the levels of quasi-continuum was investigated in Ref.~\cite{Furusaki95}; in a symmetric device ($g_1=g_2$), the maximum conductance reaches only {\it half} of the conductance quantum $G_0$, and the width of the Coulomb blockade peak scales proportionally to temperature $T$. Below we demonstrate that, on the contrary, the maximum conductance through a Majorana island ($\Delta_P\neq 0$) equals $G_0$. Thus, upon lowering the temperature below $\Delta_P$, the maximum two-terminal conductance should increase. We also show that the superconductivity modifies the off-peak conductance, which remains finite in the limit $T\rightarrow 0$. Therefore, the two-terminal conductance $G({\cal N}_g)=G({\cal N}_g,T\rightarrow 0)$ through a Majorana island varies smoothly with the dimensionless gate voltage ${\cal N}_g$. In this paper, we study the evolution of the $G({\cal N}_g)$ function with the conductances $g_i$ and  ratio  $\Delta_P/E_C$.
\\

%{\it The model.}
In the strong tunneling regime, it is convenient to use the bosonization technique which allows one to account for charging energy and superconducting pairing non-perturbatively~\cite{Lutchyn16}. Weak reflection at the junctions can be then taken into account within perturbation theory. The effective model for a proximitized nanowire in the topological regime ({\sl i.e.}, spinless p-wave superconducting state~\cite{kitaev}) in the presence of Coulomb blockade can be written as
\begin{align}
H&=H_{\rm NW}+H_{\rm C}+H_{\rm P}+H_{\rm B}\,, \label{eq:H_AH_total1}\\
H_{\rm NW}&=\frac{v}{2\pi}\int_{-\infty}^{\infty}dx \left[ (\partial_x \theta)^2+(\partial_x \phi)^2\right]\,,
\label{HNW}\\
H_C&\!=\!E_C({N}\!-\!{\mathcal N}_g)^2\!=\!E_C\left(\frac{\phi(x_2)\!-\!\phi(x_1)}{\pi}\!-\!{\mathcal N}_g\right)^2\,,
\label{HC}\\
H_P&=-\frac{\Delta_P D}{2\pi v}\int_{x_1}^{x_2} dx \cos 2\theta\,,
\label{HP}\\
H_B&=-\sum_{i=1,2} D r_i\cos 2\phi(x_i)\,.
\end{align}
Here $v$, $\Delta_P$, and $D$ are the Fermi velocity in the nanowire, induced superconducting gap, and UV cutoff energy, respectively. Charging energy $H_C$ depends on the charge transferred into the Majorana island via the two junctions, ${N}=[\phi(x_2)-\phi(x_1)]/\pi$, with the bare charging energy $E_C=e^2/2C_{\Sigma}$ being determined by the geometrical capacitance of the proximitized nanowire $C_{\Sigma}$ (including its superconducting shell). The barriers at $x_{1,2}$ are described by the reflection amplitudes $r_{1,2}$, respectively.
Here we implicitly assume that the superconducting shell renormalizes level spacing in the nanowire so that the spacing becomes negligibly small. In this respect, our model is similar to the one of Refs.~\cite{Matveev95, Furusaki95}. The term $H_P$ accounts for the superconducting proximity effect.

Let us now consider the case $E_C\gg\Delta_P$ and $r_{1,2}\ll 1 $. In this limit, term $H_C$ of Eq.~(\ref{HC}) pins the mode $\phi(x_2)-\phi(x_1)$ responsible for changing the charge of the island. Integrating out this massive mode, one obtains an effective boundary Hamiltonian~\cite{Matveev95} valid in energy band $E \ll E_C$,
\begin{align}\label{eq:B7}
\!H_B\!=\!-\sqrt{c_0 E_C D} \,
%\cdot
r(\mathcal{N}_g)\cos \left[\phi(x_2) \!+\! \phi(x_1)\!-\!\alpha(\mathcal{N}_g)\right]\,.
\end{align}
Here $\alpha(\mathcal{N}_g)$ is unimportant phase, parameter $r(\mathcal{N}_g)$ is
\begin{align}\label{eq:r}
r(\mathcal{N}_g)\!&=\!\sqrt{r_2^2+r_1^2+2 r_2 r_1 \cos(2\pi \mathcal{N}_g)}\,
%\cdot
{\mbox{sgn}}(\cos \pi \mathcal{N}_g)\,,
%\alpha(N_g)\!&=\!\tan^{-1}\left[\frac{r_1-r_2}{r_1+r_2}\tan \pi \mathcal{N}_g \right]
\end{align}
$c_0=e^C/2\pi^3$, and $C=0.5772$ is the Euler's constant. The coupling $r(\mathcal{N}_g)$ is relevant and
grows under the Renormalization Group (RG) procedure according to ${d r}/{d\ell}= {r}/{2}$ until either the running cut-off $D$ reaches $\Delta_P$ or the boundary perturbation $H_{B}$ reaches the strong-coupling limit $H_{B}\sim D$. The latter occurs at $D\sim D_c$ defined as
\begin{align}\label{eq:Gamma}
D_c \sim \Gamma_0(\mathcal{N}_g)=\frac{2e^C}{\pi^2}E_C r^2(\mathcal{N}_g),
\end{align}
where we chose the numerical coefficient in accordance with
%to fit the definition of $\Gamma_0(\mathcal{N}_g)$ in
Ref.~\cite{Furusaki95}.
The linear conductance strongly depends on the gate voltage as long as $\Gamma_0(\mathcal{N}_g)\gg \Delta_P$. In the opposite limit,
$\Delta_P\gg\Gamma_0(\mathcal{N}_g)$, conductance only weakly depends on ${\cal N}_g$,  and approaches the unitary limit.\\

%{\it Conductance away from the charge degeneracy points.}
We start by considering the limit $\Gamma_0(\mathcal{N}_g) \gg \Delta_P$ which (at sufficiently large $r_1$ and $r_2$) is realized far away from the charge degeneracy points. Upon reducing the band width $D$ in the course of RG flow to $D\sim D_c$, the combination of fields $\phi(x_1)+\phi(x_2)$ becomes pinned by the backscattering term Eq.~(\ref{eq:B7}). At smaller energy scales, $D \ll \Gamma_0(\mathcal{N}_g)$, the dynamics of $\phi(x_1)$ and $\phi(x_2)$ consists of hops between the equivalent minima of energy Eq.~(\ref{eq:B7}) which defines the two-dimensional ``landscape'' in the plane of $\phi(x_1), \phi(x_2)$. The least-irrelevant hopping term in the effective low-energy Hamiltonian shifts $\phi(x_1)+\phi(x_2)$ by $2\pi$,
\begin{align}
\!\tilde{H}_{B}\!&\!=\! -\! \lambda(D) D \cos\left[\theta (x_2^+)\!-\!\theta(x_2^-)\!+\!\theta (x_1^+)\!-\!\theta(x_1^-)\right].
\label{HB}
\end{align}
Here the fields $\theta(x_2^-,\tau)$ and $\theta(x_1^+,\tau)$ refer to the proximitized nanowire
%covered by the superconducting shell,
whereas points $x_1^-$ and $x_2^+$ belong to the leads, see Fig.~\ref{fig:device}a. At the crossover energy scale, $D\sim D_c$, the running constant $\lambda(D_c)\sim 1$, and it decreases upon reducing the band width. The RG flow for $\lambda$ in the domain $\Gamma_0({\mathcal N}_g) \gg D \gg \Delta_P$ is controlled by ${d \lambda}/{d\ell}=-\lambda$ and yields $\lambda(D)\sim \lambda(D_c) D/D_c\sim D/D_c$. The dynamics of fields $\theta(x_{1,2})$ on energy scales $E\lesssim D$ is governed by Eqs.~(\ref{HNW}), (\ref{HP}), and (\ref{HB}) with the boundary conditions $\partial_x\theta(x_{1,2}^{\pm})=0$, compatible with Eq.~(\ref{HB}).

Hamiltonian~(\ref{HB}) corresponds to an electron transfer into one end of the proximitized wire, while another electron is taken out from the opposite end. This way, a single electron charge $e$ is transferred between the leads. The corresponding current operator reads
\begin{align}
I\!=\!e\lambda(D) D \sin\!\left[\theta (x_2^+)\!-\!\theta(x_2^-)\!+\!\theta (x_1^+)\!-\!\theta(x_1^-)\right]\,.
\label{I}
\end{align}
Using it, one may evaluate the two-terminal conductance at temperatures $T\ll \Gamma_0(\mathcal{N}_g)$ by means of Kubo formula~\cite{Aleiner'02},
\begin{align}\label{eq:conductance}
G&=\frac{1}{2T}\int_{-\infty}^{\infty}dt {\Pi}\left(it +\frac{1}{2T}\right), \,  \Pi (\tau)=\langle I(\tau)I(0) \rangle
\end{align}
(here $\tau$ is imaginary time). In the intermediate range of temperatures, $\Delta_P\ll T\ll\Gamma_0(\mathcal{N}_g)$, one may ignore the pairing interaction Eq.~(\ref{HP}) and use the free-field action
%$S_{\rm NW}\{\theta(x,\tau)\}$
to determine the time evolution of the current operator Eq.~(\ref{I}). The result for the conductance $G(\mathcal{N}_g,T)$ is
\begin{equation}
\frac{G(\mathcal{N}_g,T)}{G_0}=c_1 \frac{T^2}{\Gamma_0^2(\mathcal{N}_g)}.
\label{GT2}
\end{equation}
Finding the numerical coefficient $c_1$ here is beyond the accuracy of the RG treatment, but it is known from the exact solution, $c_1 =\frac{\pi^2}{6}$~\cite{Furusaki95}.

At  lower temperatures, $T\ll\Delta_P$, fluctuations of the field $\theta (x,\tau)$ within the proximitized wire ($x_1<x<x_2$) are suppressed by the superconducting pairing term, Eq.~(\ref{HP}). To evaluate the conductance, we may reduce the band width down to $D\sim\Delta_P$, yielding $\lambda(\Delta_P)\sim\frac{\Delta_P}{D_c}$ in Eq.~(\ref{HB}), where now fields $\theta(x_1^+,\tau)$ and $\theta(x_2^-,\tau)$ are pinned to a minimum of pairing energy. With these fields being pinned, Eq.~(\ref{HB}) describes tunneling of an electron between points $x_1^-$ and $x_2^+$ belonging to the opposite leads. The corresponding tunneling action takes the form
\begin{align}
\!\!S_B\!=\!\int \!\!\frac{d\omega}{2\pi}\frac{|\omega|}{2\pi} |\theta^-|^2 \!-\!\int_{\Delta_P^{-1}}^{1/T} \!\!d
\tau \! \lambda(D) D  \!\cos \sqrt 2 \theta^-,\label{eq:weaktunneling}
\end{align}
where $\theta^-=[\theta (x_2^+)-\theta(x_1^-)]/\sqrt{2}$. Note that the boundary perturbation term in Eq.\eqref{eq:weaktunneling} becomes marginal now ({\it i.e.}, ${d \lambda}/{d\ell}=0$), and the problem at hand maps onto weak tunneling of a free fermion across an impurity. Using Kubo formula~\eqref{eq:conductance}, one can readily calculate two-terminal conductance to find
\begin{align}
\frac{G(\mathcal{N}_g)}{G_0} &=c_2\cdot\frac{\Delta_P^2}{\Gamma_0^2(\mathcal{N}_g)}\,,\quad c_2\approx \pi^2\,.
\label{GDeltaP}
\end{align}
This temperature-independent conductance is due to elastic cotunneling processes in which an electron enters the BCS condensate at one end of the wire with another electron exiting the condensate from its opposite end, leaving no excitations behind. The results obtained in the adjacent temperature intervals, Eqs.~(\ref{GT2}) and (\ref{GDeltaP}),  match each other at $T\sim\Delta_P$.

To find the numerical coefficient $c_{2}$ in Eq.~(\ref{GDeltaP}) we used the following re-fermionization procedure. First, we write the low-energy Hamiltonian~(\ref{HB}) in the fermion representation:
\begin{equation}
\tilde{H}_{B}=\frac{c_3 v^2}{\Gamma_0(\mathcal{N}_g)}
\psi^\dagger(x_2^+)\psi^\dagger(x_1^+)\psi(x_1^-)\psi(x_2^-) + {\rm h.c.}\,.
\label{HBF}
\end{equation}
Note that Eq.~(\ref{HBF}) conserves the number of electrons in the proximitized wire, so it commutes with the charging energy Eq.~\eqref{HC}. To find the numerical coefficient $c_3$, we use Eq.~(\ref{HBF}) to derive the current operator
\begin{equation}
I=i\frac{c_3 ev^2}{\Gamma_0(\mathcal{N}_g)}\left(
\psi^\dagger(x_2^+)\psi^\dagger(x_1^+)\psi(x_1^-)\psi(x_2^-) - {\rm h.c.}\right).
\label{IF}
\end{equation}
Then, applying the Kubo formula Eq.~(\ref{eq:conductance}) and ignoring the pairing interaction, we re-derive the conductance in the re-fermionized scheme and match the result with Eq.~(\ref{GT2}). That procedure fixes the coefficient $c_3=\pi$ in Eqs.~(\ref{HBF}) and (\ref{IF}). After that, we return to the evaluation of conductance at $T\ll\Delta_P$.
We are interested in the dominant, elastic contribution to the conductance; in a long proximitized wire segment, only the Majorana states contribute to the elastic amplitude~\footnote{other states would result in an inelastic transition, with two quasiparticles created in the proximitized nanowire}. The corresponding part of the Hamiltonian~\eqref{HBF} can be written in terms of Majorana fermion operators $\gamma_1$ and $\gamma_2$ localized, respectively, at $x_1$ and $x_2$,
\begin{equation}
\tilde{H}_{B}\approx-\pi v \frac{\Delta_P}{\Gamma_0(\mathcal{N}_g)}
\psi^\dagger(x_2^+)\psi(x_1^-)\gamma_1 \gamma_2 + {\rm h.c.}\,.
\label{HBF1}
\end{equation}
Finally, the calculation of the two-terminal conductance by means of the Kubo formula~\eqref{eq:conductance} applied to the low-energy tunneling processes described by Eq.~(\ref{HBF1}) yields constant $c_2$ in Eq.~(\ref{GDeltaP}). \\

We now consider weak-reflection case, $\Gamma_0(\mathcal{N}_g)\ll\Delta_P$, which is realized in a symmetric device at a gate voltage close to a charge degeneracy point, or at any gate voltage if the reflection amplitudes $r_{1,2}$ are sufficiently small (and not necessarily equal each other). At intermediate energy scale, $E_C\gg E\gg\Delta_P$, the pairing interaction Eq.~(\ref{HP}) and the boundary Hamiltonian Eq.~(\ref{eq:B7})  can be treated perturbatively. Thus, the only constraint on fluctuations of $\phi(x)$ and $\theta(x)$ within the proximitized wire is the pinning of the combination of fields $\phi(x_2)-\phi(x_1)$ by charging energy. As follows from Ref.~\cite{Furusaki95}, the conductance in the regime $T\gg \Delta_P \gg \Gamma_0$ is $G\approx G_0/2$~
\footnote{This is to be contrasted with the case of a truly one-dimensional ballistic channel where conductance is equal to $G_0$~\cite{Safi_Schulz, Maslov_Stone}}.

Upon reducing the temperature below $\Delta_P$, the pairing interaction~(\ref{HP}) suppresses the fluctuations of $\theta(x)$ within the proximitized segment, i.e. $\partial_\tau\theta(x,\tau)=0$. Thus, the condition $\partial_x\phi (x_{1,2},\tau)=0$ is enforced at the ends of the segment. To evaluate conductance in the limit of no backscattering ($\Gamma_0 (\mathcal{N}_g) \to 0$), we
integrate out modes away from $x_1$ and $x_2$ to obtain the boundary action in terms of the relevant degree of freedom $\phi^+=(\phi(x_1)+\phi(x_2))/\sqrt{2}$,
\begin{align}\label{eq:S0}
S_0=\int_0^{\Delta_P} \frac{d\omega}{2\pi} \frac{|\omega| }{2\pi}|\phi^+|^2\,.
\end{align}
The dc conductance is obtained then by using Kubo formula Eq.~\eqref{eq:conductance}; current operator in this limit is given by
\begin{align}
I=\frac{e}{2\pi} [\partial_t \phi(x_1)+\partial_t \phi(x_2)]=\frac{e}{2\pi} \sqrt 2 \partial_{t} \phi^+\,.
\end{align}
Upon evaluating  $\Pi(\tau)=\frac{e^2}{2\pi^2} \langle \partial_{\tau} \phi^+(\tau) \partial_{\tau'} \phi^+ (\tau')\rangle_{\tau'=0}$ using
%Using the correlation function
%$\langle \phi^+(\omega_n) \phi^+ (-\omega_n)\rangle=\frac{\pi }{|\omega_n|}$
%for the free-field action
Eq.~(\ref{eq:S0}), we find that $G(\mathcal{N}_g)=G_0$ in the absence of backscattering. The full quantized value of the conductance $G (\mathcal{N}_g)$ is in agreement with the notion of single-electron resonant tunneling via a Majorana state~\cite{Fu10, Heck16}. One may notice that the conductance grows by a factor of $2$ once temperature is lowered across the scale set by $\Delta_P$. This prediction can be easily verified in current experiments on proximitized nanowires~\cite{Higginbotham15,albrecht2015}.

To account for backscattering, we use Eq.~(\ref{eq:B7}) with the bandwidth $D\sim\Delta_P$,
\begin{align}
\!\!\!\!H_B\!=\!-\sqrt{c_4 E_C \Delta_P} r (\mathcal{N}_g)\cos \!\left(\phi(x_1)\!+\!\phi(x_2)\!-\! \alpha({\cal N}_g)\right)
\label{HBDelta}
\end{align}
with $c_4\sim 1$. At $E\lesssim\Delta_P$ the long-wavelength fluctuations within the proximitized wire ({\sl i.e.}, in the interval $x_1<x<x_2$) are gapped out by the pairing term~(\ref{HP}).
%At the energy scale $\sim\Delta_P$, superconducting pairing term gaps out the modes in the proximitzed nanowire (i.e. in the interval $x_1<x<x_2$).
As a result, the boundary term~\eqref{HBDelta} becomes marginal
(${dr}/{d\ell}=0$) and remains small.
% as long as $ \Gamma_0 (\mathcal{N}_g) \ll \Delta_P$.
The backscattering term Eq.~(\ref{HBDelta}) augments the free-field Hamiltonian and modifies the boundary action,
\begin{align}\label{eq:boundary}
S&=S_0- \int_{{\Delta_P}^{-1}}^{T^{-1}} d\tau \sqrt{c_4 E_C \Delta_P} \cdot r (\mathcal{N}_g)\cos (\sqrt 2 \phi^+)\,.
\end{align}
One can notice that the problem at hand maps onto the single-impurity model in the weak-backscattering limit, characterized by strong fluctuations $\phi^+$ of charge passing through the nanowire. This is to be contrasted with the strong-pinning limit, Eq.~\eqref{eq:weaktunneling}.

The evaluation of the correction to the conductance $\delta G$ within the second-order perturbation theory in  $r(\mathcal{N}_g)$, see, {\sl e.g.}, Refs.~\cite{KaneFisher3, Lutchyn2013}, yields
\begin{align}\label{eq:conductance_weak}
\frac{G(\mathcal{N}_g)-G_0}{G_0}\sim - \frac{\Gamma_0 ( \mathcal{N}_g )}{\Delta_P}\, ,
\end{align}
where $\Gamma_0 ( \mathcal{N}_g )$ is defined in Eq.~\eqref{eq:Gamma}. The numerical prefactor in Eq.~\eqref{eq:conductance_weak} is beyond the accuracy of the RG procedure. The maximal value of $\Gamma({\cal N}_g)$ equals $\Gamma_{\rm max}=(2e^C/\pi^2)E_C|r_1+r_2|^2$ and is reached at every integer ${\cal N}_g$. If the reflection amplitudes $r_{1,2}$ are small enough so that $\Gamma_{\rm max}\ll\Delta_P$, then Eq.~(\ref{eq:conductance_weak}) is applicable at all gate voltages. In the opposite case, Eq.~(\ref{eq:conductance_weak}) may be applicable in the vicinity of the half-integer values of ${\cal N}_g$, provided the setup is almost symmetric, $E_C|r_1-r_2|^2\ll\Delta_P$.\\

%{\it Evolution of the conductance oscillations: from weak to strong Coulomb blockade}.
The developed scaling theory allows us to establish the evolution of the $G({\cal N}_g)$ function upon increase of the reflection amplitudes. The two-terminal conductance oscillations with $\mathcal{N}_g$ are fully washed out by quantum fluctuations if $r_{1}$ or $r_2$ is zero. At small but finite amplitudes, $\Gamma_{\rm max}\ll\Delta_P$, oscillations are weak, see Eq.~(\ref{eq:conductance_weak}) and Fig.~\ref{fig:device}b. We will sketch further evolution of   $G({\cal N}_g)$ assuming a symmetric setup, $r_1=r_2\equiv r$. Once $r$ becomes large enough so that
$\Gamma_{\rm max}\gg\Delta_P$, the applicability of Eq.~(\ref{eq:conductance_weak}) is confined to the vicinities of the half-integer values of ${\cal N}_g$. One may use Eq.~(\ref{GDeltaP}) to estimate conductance away from these degeneracy points. %The amplitude of the conductance oscillations becomes $\sim G_0$.
Matching Eqs.~(\ref{eq:conductance_weak}) and (\ref{GDeltaP}) with each other, we find
\begin{equation}
\eta\sim [\Delta_P/E_C(1-g)]^{1/2}
\label{eq:width1}
\end{equation}
for the width of the conductance maxima, see Fig.~\ref{fig:device}b.

Further increase of the reflection amplitudes eventually results in the crossover to a weak-tunneling regime, $g_{1,2}\ll 1$. Considering it, we will still concentrate on a symmetric setup, $g_1=g_2\equiv g$. At $\Delta_P=0$, quantum fluctuations of charge of the island result in the logarithmic renormalization of the transmission amplitudes of the two junctions connecting it with the leads~\cite{Furusaki95}. Due to this ``charge Kondo'' renormalization, the transmission amplitudes reach value $\sim 1$ at the energy scale $T_K\approx E_C\exp(-\pi^2/2\sqrt{g})$, if ${\cal N}_g$ is tuned to a narrow region, $|{\cal N}_g-1/2|\lesssim T_K/E_C$. The presence of $\Delta_P$ does not prevent the aforementioned logarithmic renormalization as long as $\Delta_P\ll T_K$. At energy scales below $T_K$ we may use the strong-tunneling RG theory developed above, with the proper replacement of the parameters. Namely, in Eq.~(\ref{eq:B7}) we change $E_C\to T_K$ and modify $r({\cal N}_g)$ from the one given in Eq.~(\ref{eq:r}) to $r({\cal N}_g)\sim(E_C/T_K)({\cal N}_g-1/2)$. As the result, energy scale $\Gamma_0 ({\cal N}_g)$ of Eq.~(\ref{eq:Gamma}) is changed to $\tilde{\Gamma}_0({\cal N}_g)\sim (E_C^2/T_K)({\cal N}_g-1/2)^2$. We may use $\tilde{\Gamma}_0({\cal N}_g)$ to estimate the conductance with the help of Eqs.~(\ref{eq:conductance_weak}) and (\ref{GDeltaP}). It easy to see that the maxima of $G({\cal N}_g)$ under the considered conditions
%($\Delta_P\ll T_K$, $g\ll 1$)
have width
\begin{equation}
\eta\sim\sqrt{\Delta_P T_K/E_C^2}\,,\quad \Delta_P\lesssim T_K%\approx E_Ce^{-\pi^2/2\sqrt{g}}\,
.
\label{eq:width2}
\end{equation}
At even smaller $g$, the gap $\Delta_P$ exceeds $T_K$ and cuts off the logarithmic renormalization of the transmission amplitudes before those reach the strong-tunneling limit. As the result, $G({\cal N}_g)$ corresponds to a conventional Breit-Wigner resonance~\cite{Heck16}, with the width defined by the properly renormalized tunneling amplitudes~\cite{Furusaki95},
\begin{equation}
\eta\sim\frac{\Delta_P}{E_C}\frac{g/4\pi}{\cos^2\left[\frac{\pi}{2}\frac{\ln(E_C/\Delta_P)}{\ln(E_C/T_K)}\right]}\,,\quad \Delta_P\gtrsim T_K\,.
\label{eq:width3}
\end{equation}
Notice that Eq.~(\ref{eq:width2}), valid at an intermediate range of conductances (defined by the ratio $\Delta_P/E_C$), matches the strong- and weak-tunneling results, Eqs.~(\ref{eq:width1}) and (\ref{eq:width3}), at $T_K\sim E_C$ and $T_K\sim \Delta_P$, respectively~\footnote{The width reaches $\eta\sim\Delta_P/E_C$ at the limit of applicability of Eq.~(\ref{eq:width3})}.\\

%{\it Conclusions.}
Coulomb blockade of electron transport across a normal-state metallic island results in oscillations of the conductance $G({\cal N}_g)$ with the gate voltage. The periodicity of oscillations corresponds to the increment $e$ of the charge ${\cal N}_g$ induced on the island by the gate. Conductance accross a proximitized wire in the topologically-nontrivial superconducting state exhibits oscillations with the same period. Yet, the behavior of the function $G({\cal N}_g)$ is drastically different. This becomes especially clear in the case of a symmetric device with two identical single-channel junctions. In the normal state, $G({\cal N}_g)$ is controlled by an unstable two-channel Kondo fixed point. The conductance maxima scale linearly with temperature $T$, reaching value $G=G_0/2$ and becoming infinitely narrow in the limit $T\to 0$. On the contrary, conductance maxima in $G({\cal N}_g)$  across a proximitized wire reach value $G=G_0$ and retain finite width at $T\to 0$. We have demonstrated that the corresponding transport problem is mapped onto single-electron tunneling at any values of the bare conductance $g$ of the junctions, and found the evolution of the $G({\cal N}_g)$ with the increase of $g$ from $g\ll 1$ to $g\to 1$.

Advances in experiments with metal-semiconductor hybrids allowed recently to map out the conductance of a normal-state metallic island connected to leads by single-channel junctions~\cite{Iftikhar'15, Jezouin'16} and confirm many of the predictions of the corresponding theory~\cite{Furusaki95}. The parallel development of the proximitized nanowires~\cite{albrecht2015} set the stage for the extension of the strong-tunneling Coulomb blockade experiments into the domain of the topological superconductivity.

This work is supported by DOE contract DEFG02-08ER46482 (LG).

%\bibliography{andreevconductance1}
%merlin.mbs apsrev4-1.bst 2010-07-25 4.21a (PWD, AO, DPC) hacked
%Control: key (0)
%Control: author (8) initials jnrlst
%Control: editor formatted (1) identically to author
%Control: production of article title (-1) disabled
%Control: page (0) single
%Control: year (1) truncated
%Control: production of eprint (0) enabled
%

\end{document}